\documentstyle[11pt,aaspp4,epsfig]{article}

\slugcomment{To appear in The Astrophysical Journal Letters}
\lefthead{Cui et al.}
\righthead{RXTE OBSERVATIONS OF GRS~1737-31}

\begin{document}

\title{RXTE OBSERVATIONS OF A NEW X-RAY TRANSIENT: GRS~1737-31}
\author{Wei Cui\altaffilmark{1}, W.~A.~Heindl\altaffilmark{2}, J.~H.~Swank\altaffilmark{3}, D.~M.~Smith\altaffilmark{4}, E.~H.~Morgan\altaffilmark{1}, R~.Remillard\altaffilmark{1}, and F.~E.~Marshall\altaffilmark{3}
}
\altaffiltext{1}{Center for Space Research, Massachusetts Institute of Technology, Cambridge, MA 02139}

\altaffiltext{2}{Center for Astrophysics and Space Sciences, University of California, San Diego, La Jolla, CA 92093}

\altaffiltext{3}{NASA/Goddard Space Flight Center, Code 662, Greenbelt, MD 20771}

\altaffiltext{4}{Space Sciences Laboratory, University of California, Berkeley, Berkeley, CA 94720}
\authoremail{cui@space.mit.edu}

\begin{abstract}

We present the results from {\it RXTE} observations of a new X-ray transient, 
GRS~1737-31, near the Galactic center during an outburst. Over a span of two 
weeks, the source was seen to vary significantly on time scales down to a 
few seconds. The rapid flares in the light curves bear resemblance to 
those observed for Cygnus X-1, a well-known black hole candidate (BHC). 
The power-density spectrum is also very typical of BHCs in the {\it hard} 
(or {\it low}) state: it is flat below a characteristic break frequency 
of $\sim$0.03 Hz, and becomes roughly a 1/f power law above. The observed
X-ray spectrum can be characterized by a simple power law with a photon 
index of $\sim$1.7 over a broad energy range 2--200 keV, which is remarkably 
similar to that of Cyg X-1 in the {\it hard} state. The similarities to 
Cyg~X-1 make GRS~1737-31 a likely BHC. However, unlike most transient BHCs, 
GRS~1737-31 does not seem to reach a ``soft state'' even during an outburst.
We discuss this phenomenon in light of the comparison to a class of hard 
X--ray BHCs including transients GRO~J0422+32, V404Cyg and GRO~J1719-24, and 
persistent sources 1E1740.7-2942 and GRS~1758-258. 

\end{abstract}


\section{Discovery and ASM Light Curve}

A new X-ray transient, GRS~1737-31, was discovered during an outburst by the 
{\it SIGMA} detector aboard the {\it GRANAT} satellite on March 14, 1997 
(Sunyaev et al. 1997). The source position was later refined with follow-up 
observations with the Wide Field Camera on the {\it BeppoSAX} satellite 
(Heise 1997) and again with ASCA observations (Ueda et al. 1997).
In retrospect, a re-analysis of the scans across the region by the {\it
Proportional Counter Array} (PCA) on the {\it Rossi X-ray Timing Explorer}
(RXTE) showed that the source was in fact detected by the PCA as early as 
February 17-20, and peaked about 2 weeks later (Marshall \& Smith 1997). A 
brief {\it RXTE} pointed observation was carried out on March 21, 1997. 

The preliminary results from both the {\it GRANAT} and {\it RXTE} observations
showed that GRS~1737-31 had a hard X--ray spectrum during the outburst, which 
was very similar to Cyg~X--1 in the {\it hard} (or {\it low}) state (Sunyaev 
et al. 1997; Cui et al. 1997a). The inferred hydrogen column density suggested
a source location near the Galactic center (Cui et al. 1997a). Assuming a 
distance of 8.5 kpc, the 2-150 keV source luminosity was $1.9\times 10^{37} \mbox{ }erg\mbox{ }s^{-1}$, again similar to that of Cyg~X--1 (Cui et al. 1997b; 
Zhang et al. 1997). Similarities between the two extended further to general 
temporal properties, which suggested that GRS~1737-31 could be a BHC as well
(Cui et al. 1997a). 

The X-ray database of the All-sky Monitor (ASM) on the {\it RXTE} satellite
(Levine et al. 1996) contains scans of GRS~1737-31, although no data 
processing was carried out for this source prior to its discovery since the 
average flux was below the current threshold for new source detections near 
the Galactic center. Retrospective ASM analysis was performed, and Figure~1 
shows a light curve of the source in the ASM energy band (1.3-12 keV) in 2-day
bins (excluding data points with error bars greater than 1.0 $counts\mbox{ }s^{-1}$). The light curve shows that GRS~1737-31 started an outburst some time 
after February 14, 1997 (MJD 50493), in good agreement with the PCA scan 
results. Averaging over a period MJD 50100-50400 gives an ASM flux $\sim$7.6 
mCrab. Further pointed observations were made with the PCA and the {\it High 
Energy X--ray Timing Experiment} (HEXTE) aboard {\it RXTE}. In the following 
section, we present the results from our detailed timing and spectral analyses
of the pointed observations.

\section{Pointed RXTE Observations and Analyses}

The details of the observations are summarized in Table~1, with the pointings
slightly offset from the {\it ASCA} position in order to ensure that another
source (X1732-304) was outside of the field of view (FOV). The PCA and HEXTE 
share a common $1^{\circ}$ FOV (FWHM). The PCA (Jahoda et al. 1996) consists 
of five nearly identical large area proportional 
counter units (PCUs), with a total collecting area of about 6500 $cm^2$. 
All five PCUs were used for the first two observations. However, shortly 
after the last one started, one PCU was turned off for instrument safety
reasons. The PCA covers an energy range 2--60 keV with a moderate energy 
resolution ($\sim 18\%$ at 6 keV). The HEXTE has a total effective 
area of about 1600 $cm^2$ in two clusters. It covers a wide energy range from 
about 15 to 250 keV with an energy resolution of $\sim 16\%$ at 60 keV. The 
two clusters alternately rock on and off the source to provide nearly 
simultaneous background measurement. 

Because GRS~1737-31 is so close to the Galactic center, we were concerned 
about contamination by the ``Galactic Ridge'' emission (e.g., Worrall et 
al. 1982; Warwick et al. 1985; 1988; Koyama et al. 1989). Since this diffuse 
component is not 
included in the PCA background model, a set of off-source pointings were 
made in the immediate vicinity of the source (see Table~2 for details), in 
order to accurately model the PCA background.

\subsection{Timing Analysis}

Only the PCA data are used here. GRS~1737-31 displayed large X--ray 
variability with occasional flares that lasted a few seconds. Figure~2 shows 
a typical example of the light curve, made from the ``Standard1'' data with 
a 1/8 s timing resolution. 

High timing resolution data ($\sim 1\mu s$) are available through two 
``goodXenon'' modes. Using the {\it Fast Fourier Transform} (FFT) technique, 
a search for high-frequency features (e.g., quasi-periodic oscillations) was 
carried out at frequencies up to a few kiloHertz, but none was detected. In 
fact, most of the power lies below a few Hertz. For a better low-frequency 
coverage, we broke up the ``Standard1'' data into 256-s segments for each 
observation, performed FFT on each segment, and averaged the results. 
Figure~3 shows the average power-density spectra (PDS). Note that the power 
density is expressed in terms of the fractional rms amplitude squared, which 
is defined as Leahy normalized power (with Poisson noise power subtracted) 
divided by the mean source count rate (van der Klis 1995). Instrumental 
artifacts due to electronic deadtime and very high energy events (Zhang et al.
1996) were taken into account.

There is no apparent temporal evolution between the observations: the PDS is 
flat below $\sim$0.03 Hz, and drops roughly as a simple 1/f power law above. 
Integrating over a range 0--4 Hz, the fractional rms variability is about 30\%
for all observations. It is larger than our preliminary estimate (Cui et al. 
1997a) because we are now including the ``Galactic Ridge'' emission in the 
background modeling. 

\subsection{Spectral Analysis}

On-source time intervals were selected from the raw light curves, and were 
used to extract an X-ray spectrum for each observation. We simultaneously
fit the individual PCU spectra and HEXTE cluster spectra with a simple power 
law plus a Gaussian component (to model a line feature at $6.6-6.7$ keV 
which is apparent in the raw spectra). The relative normalization between 
the PCA and HEXTE was allowed to vary due to the uncertainty in their relative 
effective areas. The results are summarized in Table~3. 

In fitting the PCA and HEXTE spectra, we alternately used the individual
PCA off-source pointings as well as the average result to represent the
PCA background spectrum. The variation in the background ($<$ 20\% in total
count rate) provides a realistic estimate of the systematic uncertainty for 
the analysis, since this effect dominates the statistical and any PCA 
calibration uncertainties. The ``best-fit'' parameters were derived by 
using the average off-source spectrum as the PCA background, and the 
uncertainties represent the range of each parameter derived using different 
off-source spectra. 

The simple model characterizes the observed X-ray spectrum reasonably well 
over a broad energy range 2-200 keV, as indicated by the small reduced 
$\chi^2$ values (which are based on only the statistical uncertainty of the 
data). As an example, Figure~4 shows the
spectrum from Observation~2, with the average off-source spectrum subtracted, 
along with the best-fit model and residual spectrum. Although no significant
emission is apparent above $\sim$130 keV from the figure, the source can be 
detected up to $\sim$200 keV when combining data from all observations. 

To investigate the presence of a possible ultra-soft component, typical for 
BHCs, we added a black body component to the model, and re-fitted the data.
Only in some of the cases (where different off-source spectra were used for 
the PCA background) was a non-zero normalization for this component obtained, 
providing slightly improved fits. In those cases, the black body temperature 
is in the range 0.9-1.5 keV, and the soft component accounts for less than 
$\sim$10\% of the intrinsic 2-10 keV flux. Neither was an ultra-soft component
needed to fit the ASCA data (Ueda et al. 1997).

\section{Discussion}

Both the light curves and PDS of GRS~1737-31 during the outburst are 
remarkably similar to Cyg~X-1 in the {\it hard} state (see review by van der 
Klis 1995, and references therein). Recent works on Cyg~X-1 (Cui et al. 
1997b,c,d) showed that the white noise (i.e., the flat portion at low 
frequency) may be intrinsic to the 
source, originating in the statistical fluctuation associated with the 
accretion process near the black hole where the dynamical time scale is very
short. It was speculated that the intrinsic PDS is then modified by the 
scattering process occurring in a hot electron corona which acts as a low-pass 
filter. The break frequency is simply determined by the characteristic photon 
escape time from the corona (thus its physical size). If GRS~1737-31 is 
also a black hole binary, the observed break frequency would imply a similar 
corona size to that of Cyg X-1 in the hard state.

The observed X-ray continuum of GRS~1737-31 can be characterized by a simple 
power law with a photon index of $\sim$1.7 over an energy range of 2-200 keV. 
The power-law spectrum is very typical of BHCs, and used to be considered as 
one of their defining signatures. In fact, this photon index is also typical 
for Cyg X-1 in the hard state. For Cyg X-1, the power-law spectrum is thought 
to be the product of inverse Comptonization of the soft photons in the region 
by the hot corona (e.g., Shapiro, Lightman, \& Eardley 1976). Similar physical
processes may also be involved in GRS~1737-31. However, unlike most transient
BHCs, GRS~1737-31 does not seem to reach a ``soft state'' even during an 
outburst. This phenomenon was also observed for persistent BHCs 1E1740.7-2942 
and GRS~1758-258.

One of the important spectral characteristics for a typical BHC is the presence
of a prominent ultra-soft component, which is thought to be the emission from 
the inner edge of the accretion disk. It has been a long standing puzzle in 
black hole studies why such a component has {\it never} been seen for some
of the known BHCs, such as transients V404Cyg, GRO~J0422+32, and GRO~J1719-24,
and persistent sources 1E1740.7-2942 and GRS~1758-258. The lack of a strong 
soft component seems to indicate that GRS~1737-31 is perhaps another one of 
this type. Recently, Zhang, Cui, and Chen (1997) argued that the strength 
of the ultra-soft component may be directly 
related to the spin of the black hole, which determines the location of the 
last stable orbit, since there are no known processes to keep the accretion 
disk from extending all the way to the last stable orbit in the high state. 
They speculated that the absence of an ultra-soft component may indicate
the presence of a rapidly rotating black hole in the system with a retro-grade
disk. In such cases, the last stable orbit, thus the inner edge of the disk, 
is farther away from the black hole than the non-rotating ones, so the 
temperature of the region may be too low for the emission to be detected by 
current experiments. 

The inferred hydrogen column density for GRS~1737-31 is in a range 
4.8--7.0$\times 10^{22}\mbox{ }cm^{-2}$, in excellent agreement with the
results from the ASCA observation (Ueda et al. 1997). Therefore, GRS~1737-31 
is likely to be near the Galactic center, again similar to 1E1740.7-2942 and 
GRS~1758-258. Unlike in those cases, however, no radio or infrared 
counterparts have been identified for GRS~1737-31.

The ``Galactic Ridge'' emission is known to show strong iron emission lines 
at $\sim$6.7 keV (e.g., Koyama et al. 1986; Koyama 1989; Yamauchi,et al. 1990;
Koyama et al. 1996). Imperfect background subtraction could lead to the
appearance of an iron line in the net spectrum. We estimated that the line 
fluxes in the one degree fields of view were roughly $6.8$, $4.6$, and 
$5.5 \times 10^{-4}\mbox{ }photons\mbox{ }s^{-1}\mbox{ }cm^{-2}$, 
respectively, from the PCA background fields 1, 2, and 3. These are consistent
with the ASCA measurements (Sakano \& Koyama, private communication), using 
fields much closer to GRS~1737-31 from the image (Ueda et al. 1997).
Therefore the PCA off-source pointings seem to provide a good estimate of the
local diffuse flux. The source spectrum obtained with the PCA appears to 
contain residual iron line emission. The equivalent widths in Table~3
corresponds to line fluxes of $2.9^{+1.0}_{-1.2}$, $3.0^{+1.2}_{-1.1}$, and
$2.5^{+1.0}_{-1.0}\times 10^{-4}\mbox{ }photons\mbox{ }s^{-1}\mbox{ }cm^{-2}$.
The line could be due to the fluorescent emission from highly ionized iron 
atoms associated with the source. However, this detection should be taken
with reservation, because of the uncertainty in the PCA calibration.

\acknowledgments
We gratefully acknowledge the effort by the staff at the RXTE Science 
Operation Center for making flexible mission operation a reality. This work 
is supported in part by NASA Contracts NAS~5-30612 and NAS~5-30720.

\newpage
\begin{deluxetable}{lcrr}
\tablecolumns{4}
\tablewidth{0pc}
\tablecaption{{\it RXTE} Observations of GRS~1737-31\tablenotemark{1}}
\tablehead{
 & & \multicolumn{2}{c}{Live Time (s)} \\
\cline{3-4}
\colhead{Obs.} & \colhead{Time (UT)} & \colhead{PCA} & \colhead{HEXTE\tablenotemark{2}}
}
\startdata
1 &03/21/97 16:36:00--17:37:00 & 3248 & 1972 \nl
2 &03/28/97 12:16:00--16:04:00 & 5440 & 2571 \nl
3 &04/02/97 14:44:00--20:01:00 &10640 & 5966 \nl
\tablenotetext{1}{Pointing coordinates: $R.A.=17^h40^m24^s$, $Decl.=-31\arcdeg 00\arcmin 00\arcsec$, epoch J2000.}
\tablenotetext{2}{Both clusters are included.}
\enddata
\end{deluxetable}

\begin{deluxetable}{lcccc}
\tablecolumns{5}
\tablewidth{0pc}
\tablecaption{PCA Background Observations on 03/28/97}
\tablehead{
\colhead{Obs.} & \colhead{R.A.\tablenotemark{1}} & \colhead{Decl.\tablenotemark{1}} & \colhead{Time (UT)} & \colhead{Exposure (s)}
}
\startdata
1 &266.20\arcdeg & -31.00\arcdeg & 13:28:00--13:39:28 &608 \nl
2 &265.00\arcdeg & -33.00\arcdeg & 13:39:29--13:51:59 &736 \nl
3 &264.10\arcdeg & -32.10\arcdeg & 13:52:00--14:02:59 &576 \nl
\enddata
\tablenotetext{1}{Epoch J2000.}
\end{deluxetable}

\begin{deluxetable}{ccccccccc}
\tablecolumns{9}
\tablewidth{0pc}
\tablecaption{Results of Spectral Analysis\tablenotemark{1}}
\tablehead{
\colhead{Obs.}&\colhead{$N_H$\tablenotemark{2}}& \colhead{$\alpha$}&\colhead{$N_{\alpha}$\tablenotemark{3}}&\colhead{$E_c$} &\colhead{$E_{\sigma}$}&\colhead{E.W.}&\colhead{Flux\tablenotemark{4}}&\colhead{$\chi^2_{\nu}/dof$}\\
 & & & &(keV) & (keV) & (eV) & &
}
\startdata
1 & $5.6^{+0.4}_{-0.8}$ & $1.69^{+0.03}_{-0.07}$ & $1.19^{+0.07}_{-0.18}$ & $6.61^{+0.00}_{-0.01}$ & $0.25^{+0.29}_{-0.08}$ & $61^{+24}_{-20}$ & $2.3^{+0.3}_{-0.2}$ & $1.3^{+0.3}_{-0.2}/488$ \nl
2 & $5.5^{+0.6}_{-1.1}$ & $1.68^{+0.05}_{-0.09}$ & $0.94^{+0.07}_{-0.18}$ & $6.62^{+0.00}_{-0.02}$ & $0.3^{+0.6}_{-0.1}$ & $83^{+39}_{-29}$ & $1.8^{+0.3}_{-0.1}$ & $1.2^{+0.2}_{-0.1}/488$ \nl
3 & $6.3^{+0.7}_{-1.1}$ & $1.76^{+0.05}_{-0.10}$ & $0.89^{+0.07}_{-0.18}$ &$6.7^{+0.0}_{-0.1}$ & $0.3^{+0.4}_{-0.1}$ & $88^{+27}_{-36}$ & $1.3^{+0.2}_{-0.2}$ & $1.5^{+0.4}_{-0.4}/408$ \nl
\tablenotetext{1}{Errors represent the range of parameters derived by using 
individual off-source pointings as the background. }
\tablenotetext{2}{H I column density along the line-of-sight, in units of $10^{22}\mbox{ }cm^{-2}$.}
\tablenotetext{3}{Power-law normalization, in units of $10^{-1}\mbox{ }photons\mbox{ }cm^{-2}\mbox{ }s^{-1}\mbox{ }keV^{-1}$ at 1 keV.} 
\tablenotetext{4}{Observed 2-200 keV flux, in units of $10^{-9}\mbox{ }erg\mbox{ }cm^{-2}\mbox{ }s^{-1}$.} 
\enddata
\end{deluxetable}

\newpage
\begin{figure}
\psfig{figure=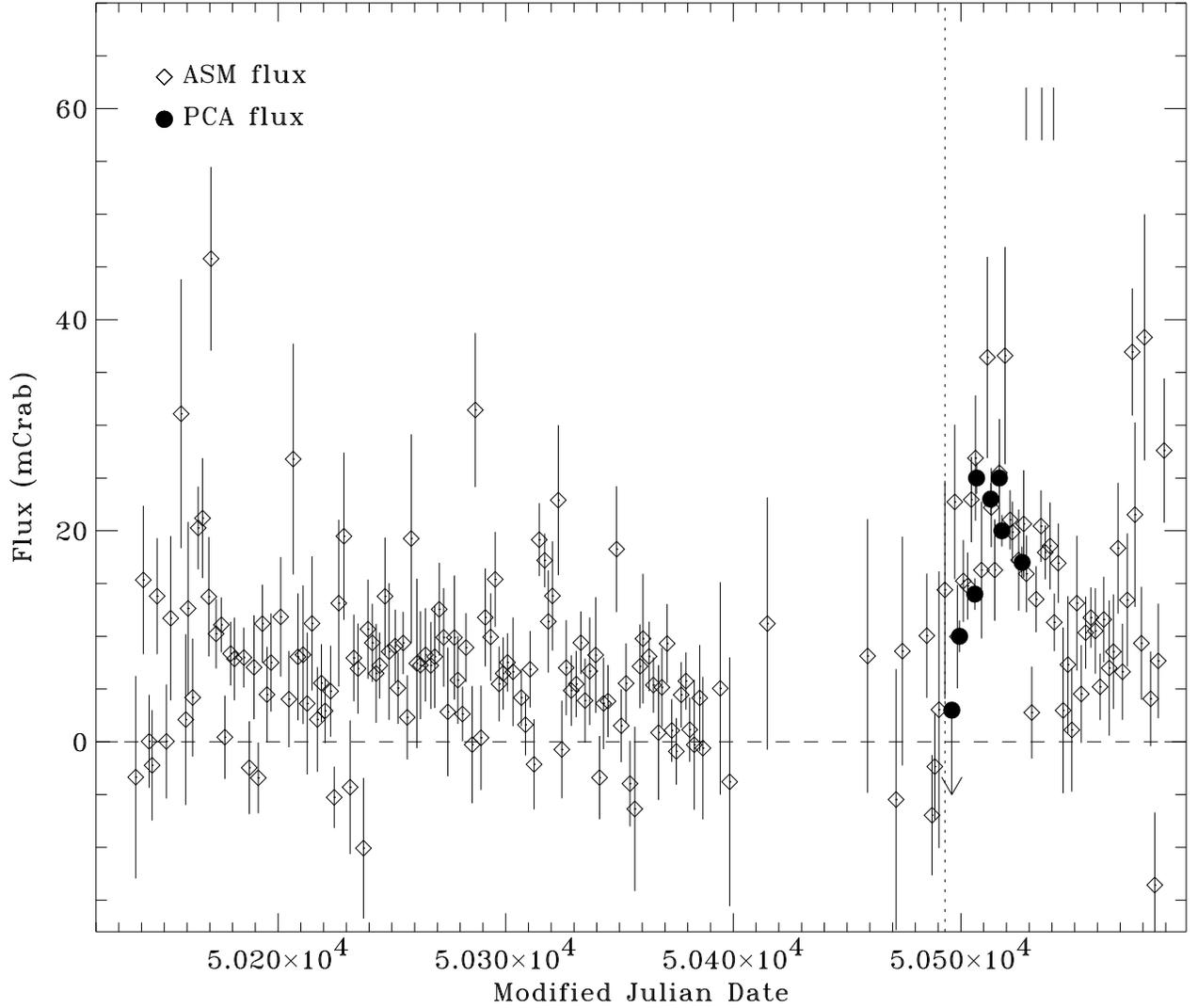}
\caption{ASM light curve of GRS~1737-31 in 2-day bins. One crab corresponds 
to about 75 $counts\mbox{ }s^{-1}$ in the ASM band (1.3-12 keV). The dotted 
line shows roughly when the outburst started. Also shown are the PCA slew 
measurements in the energy range 2-10 keV (see text), with the $1\sigma$ 
error being conservatively estimated to be 1.5 mCrab, which is dominated by 
the uncertainty in the background. The The ticks near the top indicate when 
the pointed RXTE observations were carried out. }
\end{figure}

\newpage
\begin{figure}
\psfig{figure=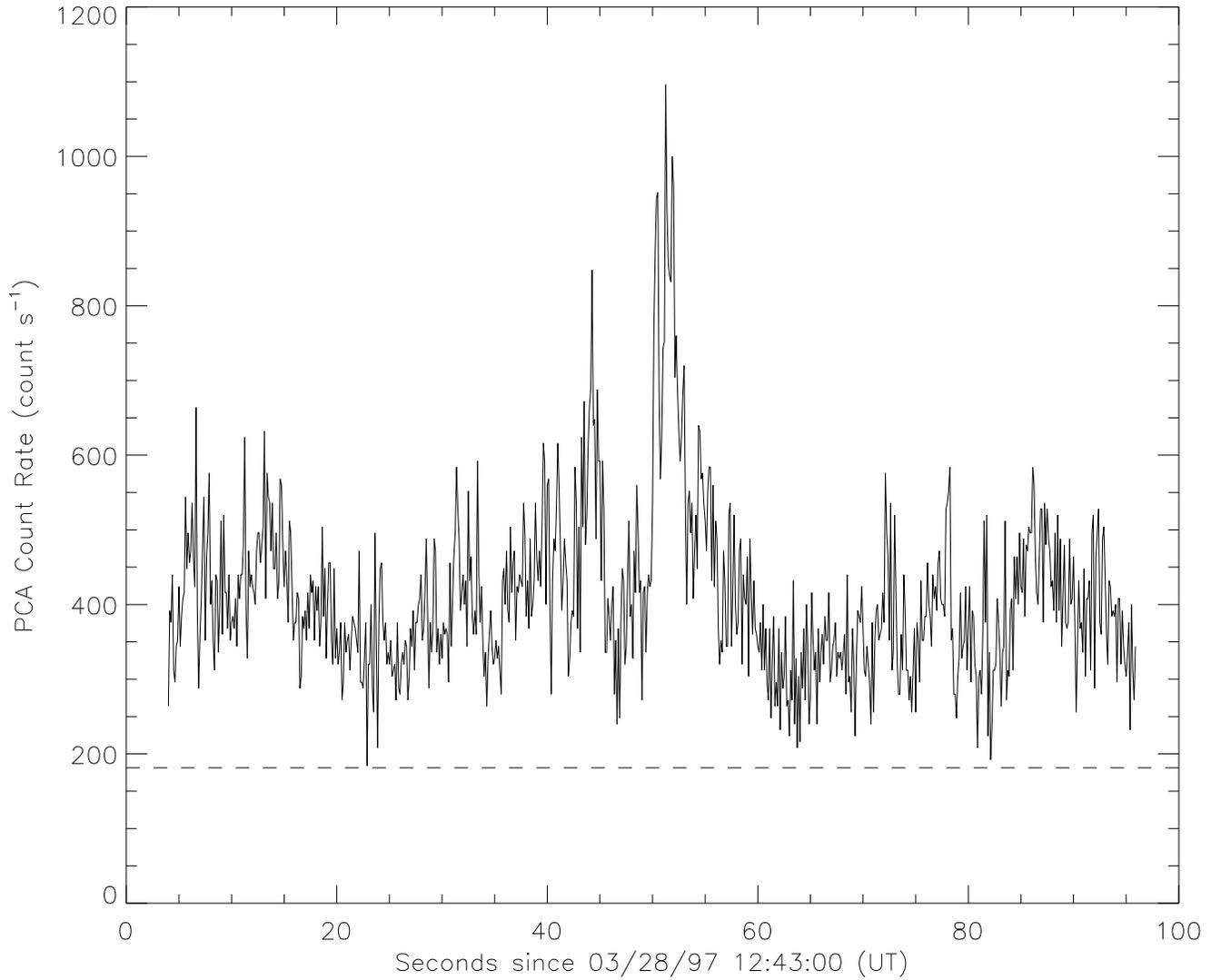}
\caption{A segment of the PCA light curve of GRS~1737-31 from the second 
observation (see Table~1). The time bin size is 0.125 s. The dashed-line 
indicates the average background count rate. }
\end{figure}

\newpage
\begin{figure}
\psfig{figure=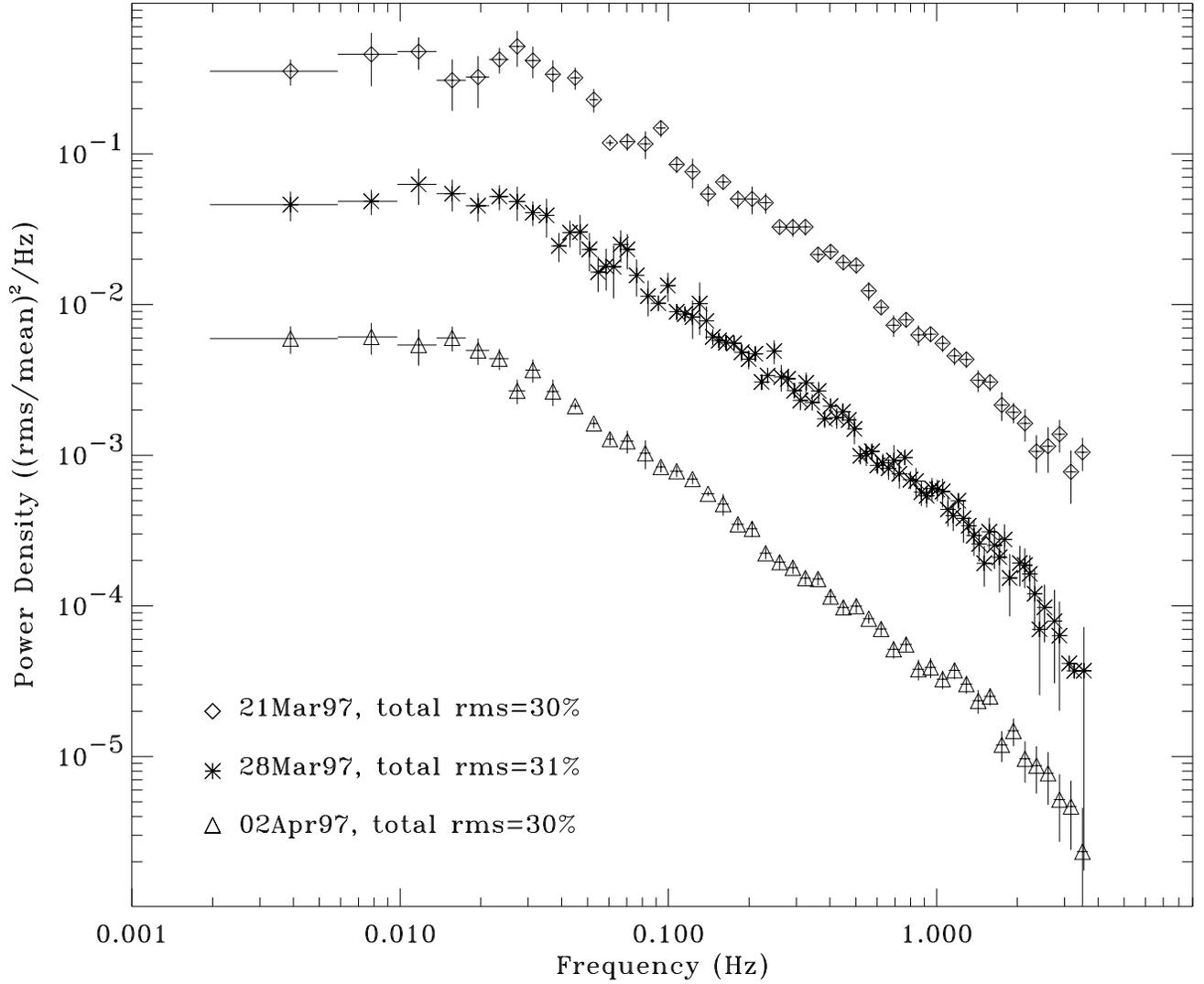}
\caption{Power density spectra of GRS~1737-31. They are derived over the
entire PCA band. For clarity, the lower two curves have been shifted down 
by a factor by 10 and 100 respectively.}
\end{figure}

\newpage
\begin{figure}
\psfig{figure=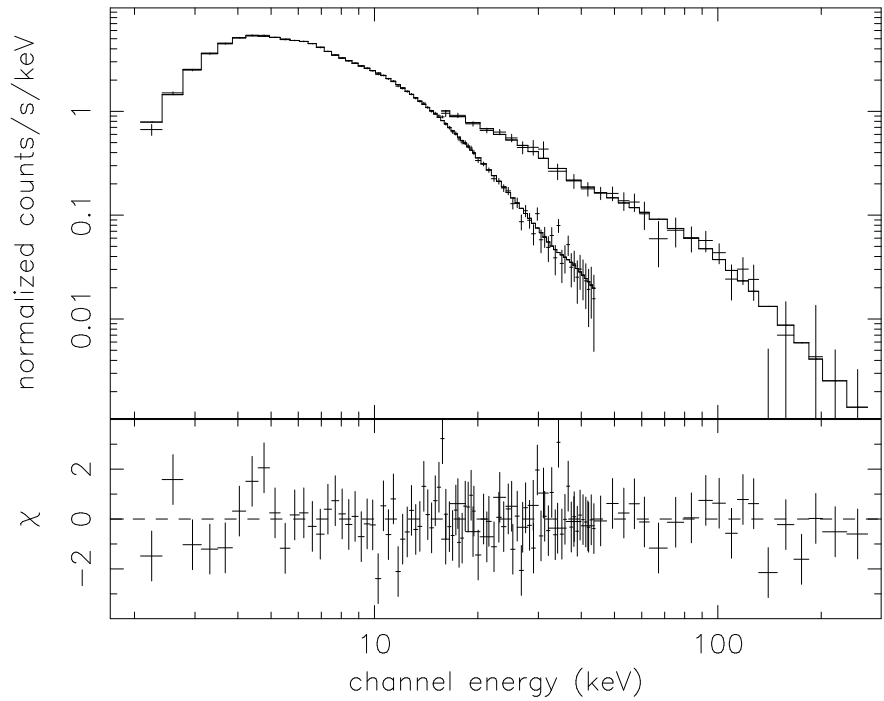,width=6.0in,angle=-90}
\caption{Joint PCA/HEXTE spectrum of GRS~1737-31 obtained from the second
observation. Note that the average off-source spectrum is subtracted as 
the PCA background (see text). For clarity, five PCU spectra are co-added,
so are the HEXTE cluster spectra. The errors shown are purely statistical.
The solid histogram shows the best-fit model.}
\end{figure}

\end{document}